\documentclass{JAC2000}

\usepackage{graphicx}

\begin{document}
\title{RELATIVISTIC ELECTRON FACILITY FOR EDUCATION AND RESEARCH AT HIROSHIMA UNIVERSITY\thanks{Work supported by the Venture Business Laboratory and the Grant in Aid for Scientific Research Monbusho No. 06452072.}}

\author{T. Ohgaki, I. Endo, M. Andreyashikin, K. Chouffani, K. Goto, S. Masuda, K. Matsukado, \\ T. Takahashi, K. Yoshida, Hiroshima University}

\maketitle

\begin{abstract}

 The Relativistic Electron Facility for Education and Research (REFER) at
 Hiroshima University accepts a 150~MeV electron beam from a microtron,
 the injector to the 700~MeV storage ring of the Hiroshima Synchrotron
 Research Center, and keep it circulating for 2.5ms without a RF
 acceleration. It acts as a beam stretcher and has been used for the
 following researches: 1) X-ray generation from an internal multiple
 foil crystalline target due to the parametric X-radiation (PXR) and the
 resonant transition radiation (RTR). 2) Laser backward Compton
 scattering. 3) Coherent pair creation from Si crystal by 150~MeV
 bremsstrahlung. 4) Study of the Ter-Mikaelian effect in the low energy
 part of the bremsstrahlung in an extracted electron line. 5) A novel
 scheme of beam stabilization by the induction coil. This paper reports
 the experimental results and the current status of the REFER electron
 ring.

\end{abstract}

\section{INTRODUCTION}

 There has been a growing interest in applications of relativistic
 electron beam from small devices to industrial~\cite{ube91}, 
 medical~\cite{wie94}, and scientific~\cite{car94} fields in recent years.

 The REFER electron ring is a compact electron circulating ring for
 application research of the relativistic electron beam and for education
 of beam physics~\cite{ref00}. The device was installed at Venture Business
 Laboratory, Hiroshima University in 1997. The electron beam energy is
 150~MeV, which is generated by the microtron~\cite{hor91} at Hiroshima 
 Synchrotron Radiation Center~\cite{yos98}. A beam extraction line is 
 attached to the REFER electron ring. The electron beam can be slowly 
 extracted from the main ring. The electron beam at the REFER is
 utilized for investigation and education of the beam physics,
 development of new X-ray sources such as PXR, RTR generation, and Laser 
 backward Compton scattering, and study of the particle physics such as
 an experiment of the coherent pair creation, etc.

 Detail of the REFER electron ring is described in the next
 section. Section 3 is devoted to the presentation of the experimental
 results at the REFER. Last section summarizes the current status of the
 REFER.   

\begin{figure}[htb]
\centering
\includegraphics*[width=72mm]{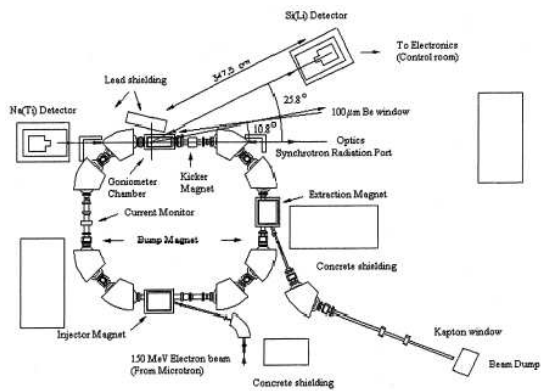}
\caption{REFER electron ring.}
\label{refer}
\end{figure}

\section{REFER ELECTRON RING}

 The 150~MeV electron beam generated by the microtron at Hiroshima
 Synchrotron Radiation Center is injected to the REFER electron
 ring. Figure~1 shows the layout of the REFER electron ring. The
 injection beam has typically 1 $\mu$s in bunch length, peak current
 upto 10~mA and pulse repetition rate of 2-100~Hz. Multi-turn injection
 into the REFER electron ring is performed by the injection septum magnet
 and the bump magnets. Pulse power generators supply pulse current to
 coils of those magnets synchronous with beam injection timing. The
 injection septum magnet puts the beam injected from microtron into the
 bump orbit. Deviation of the bump orbit from the reference orbit is
 varied from 30~mm to zero with time in 1~$\mu$s to avoid collisions of
 circulating electrons to the injection septum magnet.

 As the present REFER electron ring has no acceleration mechanism, the
 energy loss of electron beam, 59.7 eV per turn, due to synchrotron
 radiation is not compensated. Deviation of the electron orbit from the
 reference orbit increases with time. Therefore, the electrons are lost
 at vacuum pipes after several ten thousands turns. In summary, the
 REFER acts as a beam stretcher and has been used for the several
 researches described in next section last year.

\begin{figure}[htb]
\centering
\includegraphics*[width=80mm]{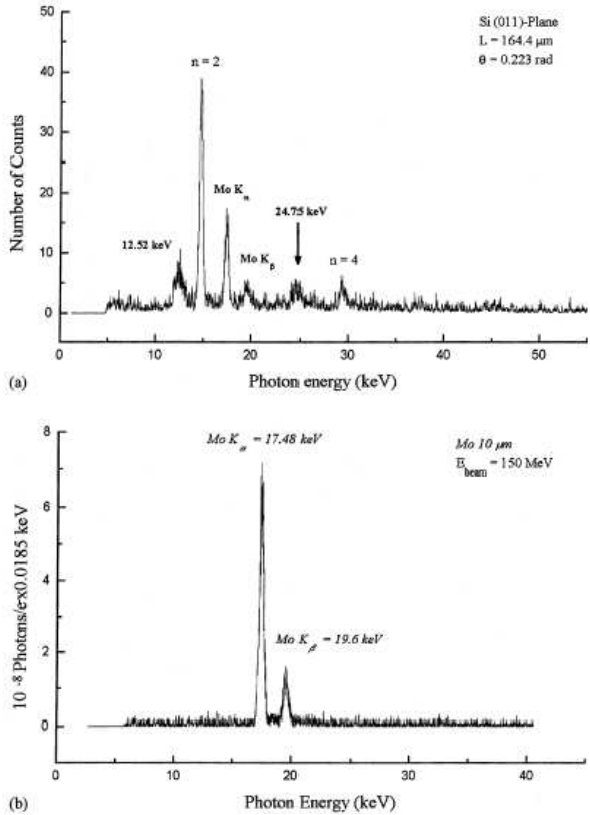}
\caption{(a) PXR from (011) plane in 164.4 $\mu$m silicon crystal with the amorphous Mo target. (b) Spectrum from the amorphous Mo target~\cite{cho00}.}
\label{pxr}
\end{figure}

\section{RESEARCH AT REFER}

\subsection{Parametric X-Radiation}

 X-radiation generated by electrons passing through silicon crystal 
 targets of different thicknesses (49.2, 164.4 and 1644.0~$\mu$m) and 
 accurately aligned 10 and 100 layers of 16.4 $\mu$m thick 
 monocrystalline silicon foils was measured at the REFER~\cite{cho01,cho00}. 
 A clear intensity enhancement was observed, when compared to the intensity
 of PXR from single crystal targets of equivalent thickness at 14.4 and 
 28.8 keV. K.~Chouffani~{\it et al.} believed that this enhancement 
 results from the diffraction of transition radiation from individual 
 surfaces of the foils off the crystallographic planes~\cite{cho01}. When 
 the molybdenum foil was used for electron beam normalization, they observed 
 an unidentified peak at 12.5 keV and its second harmonic. Figure 2 (a) 
 shows PXR spectrum from 164.4 $\mu$m silicon single crystal with the 
 amorphous molybdenum foil. We plan to investigate the unidentified
 peaks using the electron beams from the REFER extraction line in 2001.

\begin{figure}[htb]
\centering
\includegraphics*[width=90mm]{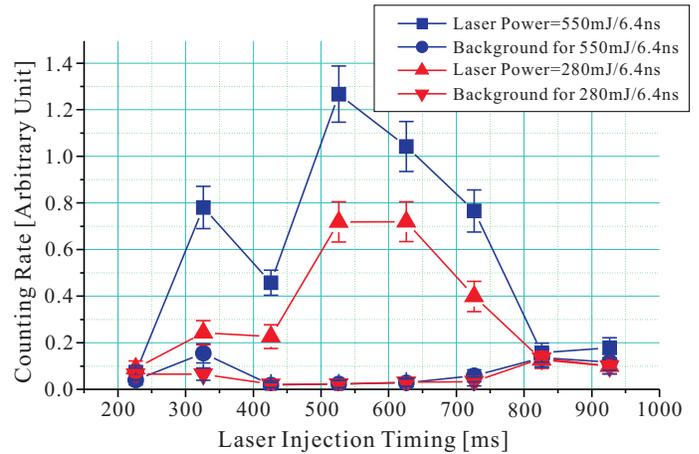}
\caption{Counting rate of the Laser backward Compton scattering versus Laser injection timing.}
\label{lbs}
\end{figure}

\subsection{Laser Compton Scattering}

 The Laser backward Compton scattering was studied at the REFER. The 
 wavelength and pulse energy of the Nd:YAG Laser beam were 532~nm and 
 550~mJ/pulse, respectively. Figure 3 shows the counting rate of the Laser 
 backward Compton scattering. The electron beam shifts to inside direction 
 of the ring and the counting rate changes with the Laser injection timing.
 The intensity and energy of the generated X-ray were about 0.1 photons 
 per bunch and less than 800~keV, respectively. This is first
 measurement of the Laser Compton scattering at the REFER and we plan to
 measure its spectrum in 2001. 

\begin{figure}[htb]
\centering
\includegraphics*[width=80mm]{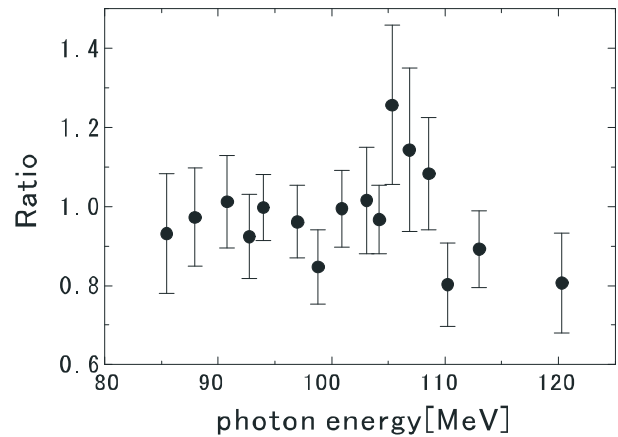}
\caption{Ratio of the pair cross section with the Si target to with the Al target~\cite{oka01}.}
\label{cpc}
\end{figure}

\subsection{Coherent Pair Creation}

 The coherent pair creation from Si crystal by 150~MeV bremsstrahlung was 
 measured at the REFER~\cite{oka01}. If a target is a single crystal, due 
 to the periodicity of the nuclear potential, the pair cross section has 
 several sudden increase against non-single crystal. The ratio of the pair 
 cross section with the Si target to that with the Al target is shown in 
 Fig.~4. Y.~Okazaki {\it et al.} observed a peak at 105~MeV in a (111) 
 Si crystal to investigate the interference effect among atomic string. 
 As a result, the statistical quantity was not sufficiency, however, a 
 gentle enhancement was observed at 105~MeV.

\subsection{Study of Ter-Mikaelian Effect}

 The Ter-Mikaelian effect in the low energy part of the bremsstrahlung 
 in an extracted electron line has been studied. It is well known that 
 the cross section of the bremsstrahlung from high-energy electron beam 
 with a single atom diverges in low energy region. According to 
 Ter-Mikaelian's theory, the effects of the multiple scattering and the 
 polarization cause the reduction of the cross section of the bremsstrahlung 
 in a medium. Using the electron beam with 150MeV energy, which has a 
 negligible effect of the multiple scattering, T.~Ohnishi {\it et al.} 
 plan to measure the polarization effect of the bremsstrahlung.

\subsection{Beam Stabilization by Induction Coil}
 
 A novel scheme of beam stabilization by the induction coil is being 
 studied~\cite{mat00}. The electron beam in the REFER loses its energy 
 by 59.7 eV per turn due to synchrotron radiation. As the REFER has not 
 acceleration instruments, the beam orbit moves toward the inner wall of 
 the vacuum pipe. It is about 2.5 ms from the injection to the dumping. If 
 the energy loss of the beam is compensated by induction acceleration, the 
 beam can keep circulating for longer time. S~Matsuno {\it et al.} designed 
 a small-sized model of an induction magnet, which has 1/8 cross section of 
 the iron core to be installed. Combining the software for the current form 
 and the small-sized model, the magnetic permeability of the core material 
 was measured to determine the most suitable form.

\begin{table}[htb]
\begin{center}
\caption{REFER experiments in 2001}
\begin{tabular}{l}
\hline
Beam Survey from REFER Extraction Line \\
Scintillation Counter \\
Parametric X-ray Radiation \\
Laser Backward Comtopn Scattering \\ 
Study of Ter-Mikaelian Effect \\
Radiated Photon from Photonic Crystal\\ \hline
\end{tabular}
\label{refer}
\end{center}
\end{table}

\section{SUMMARY}

 The outline and the current status of the REFER electron ring were
 described. In 2000 the REFER has been used for the following
 researches:
\begin{enumerate}
  \item X-ray generation from an internal multiple foil crystalline
	target due to the PXR and the RTR. 
  \item Laser backward Compton scattering. 
  \item Coherent pair creation from Si crystal by 150~MeV
	bremsstrahlung.
  \item Study of the Ter-Mikaelian effect in the low energy part of the
	bremsstrahlung in an extracted electron line. 
  \item A novel scheme of beam stabilization by the induction coil. 
\end{enumerate}
  Finally, the REFER experiments planed in 2001 are listed in Table~1.

%\begin{figure*}[t]
%\centering
%\includegraphics*[width=150mm]{JACpic2.eps}
%\caption{Example of full width figure, showing the distribution of problems commonly
%  encountered  during the processing of EPAC'98 papers.}
%\label{l2ea4-f2}
%\end{figure*}

%\begin{table}[htb]
%\begin{center}
%\caption{Margin specifications}
%\begin{tabular}{|l|c|c|}
%\hline
%\textbf{Margin} & \textbf{A4 Paper} & \textbf{US Letter Paper} \\ \hline
%Left & 20 mm & 20 mm (0.79 in) \\
%Right & 20 mm & 26 mm (1.0 in) \\
%Top & 37 mm & 19 mm (0.75 in) \\
%Bottom & 19 mm & 19 mm (0.75 in) \\ \hline
%\end{tabular}
%\label{l2ea4-t1}
%\end{center}
%\end{table}


\begin{thebibliography}{9}

\bibitem{ube91} H.~Uberall {\it et al.}, in {\it SPIE Proc. Short Wave 
        Radiation Sources 1552} (1991) 198.

\bibitem{wie94} H.~Wiedemann {\it et al.}, Nucl. Instrum. and Meth. in 
        Phys. Res. A347 (1994) 515.

\bibitem{car94} R.~Carr, Nucl. Instrum. and Meth. in Phys. Res. A347 
        (1994) 510.

\bibitem{ref00} S.~Masuda, I.~Endo, T.~Takahashi, K.~Matsukado,
	Y.~Okazaki, K.~Ikematsu, T.~Ohnishi, S.~Matsuno, D.~Iseki,
	M.~Kimura, K.~Nakamura, and A.~Sharafutdinov, in {\it
	Proceedings of the 25th Linear Accelerator Meeting}, July 12-14,
	(Himeji, Japan) 2000.

\bibitem{hor91} T.~Hori {\it et al.}, in {\it Proceedings of PAC91}, 1991.

\bibitem{yos98} K.~Yoshida {\it et al.}, in {\it Proceedings of	APAC98},
	Mar 23-27, (Tsukuba, Japan) 1998.

\bibitem{cho01} K.~Chouffani, M.Yu.~Andreyashkin, I.~Endo, J.~Masuda,
	T.~Takahashi, and Y.~Takashima, Nucl. Instrum. and Meth. in
	Phys. Res. B173 (2001) 241. 

\bibitem{cho00} K.~Chouffani, M.Yu.~Andreyashkin, I.~Endo, K.~Goto,
	J.~Masuda, T.~Takahashi, Y.~Takashima, and K.~Yoshida, in {\it
	Proceedings of the 4th Hiroshima International Symposium on
	Synchrotron Radiation},	Mar. 16-17, (Hiroshima University,
	Japan) 2000.

\bibitem{oka01} Y.~Okazaki, D.~Iseki, I.~Endo, and T.~Takahashi, The
	56th Spring meeting of Physical Society of Japan, (Chuo
	University, Japan) 2001.

\bibitem{mat00} S.~Matsuno, G.~Chakhlov, I.~Endo, and S.~Masuda, The
	meeting of Beam	Physics 2000, Dec. 11-13, (SPring-8, Japan) 2000.
 
\end{thebibliography}
\end{document}